\documentstyle[12pt]{article}
\newcommand{\be}{\begin{equation}}
\newcommand{\ee}{\end{equation}}

\newcommand{\bea}{\begin{eqnarray}}
\newcommand{\eea}{\end{eqnarray}}
\newcommand{\unop}{1\!{\rm I}}

\title{Quasi Exactly Solvable N$\times$N-Matrix  Schr\"odinger
Operators.}

\author{{\large Y. Brihaye} \\
{\small Department of Mathematical Physics, University of Mons-Hainaut}\\
{\small Place du Parc, B-7000 Mons, Belgium }\\
   {\large B. Hartmann}\\
{\small Fachbereich Physics , 
University of Oldenburg}\\
{\small Postfach 2503, D-26111 Oldenburg,Germany}}
\date{\today}
\begin{document}
\maketitle
\thispagestyle{empty}
\begin{abstract}
\par New examples of matrix quasi exactly solvable
Schr\"odinger operators are
constructed. One of them constitutes a  matrix
generalization  of the quasi exactly solvable anharmonic oscillator,
the corresponding invariant vector space is constructed explicitely.
Also investigated are matrix generalizations of the Lam\'e equation. 
\end{abstract}
\section{Introduction}
The topic of Quasi Exactly Solvable (QES) operators \cite{tur1,ush}
has been developped in the last years. It consists of differential
operators (mainly Schr\"odinger ones) which possess a finite dimensional
invariant vector space, say ${\cal V}$, of functions. So, the
restriction of the eigenvalue equation to the space ${\cal V}$
leads to an algebraic eigenvalue problem.

Although scalar QES operators have been classified in one \cite{tur2}
and several \cite{olver} variables, a classification of matrix 
QES operators is still missing. Yet, interesting examples of them 
have been obtained in relation with the stability analysis of 
soliton solutions occuring in some field theories \cite{bk1,bt}.    

This problem was first addressed in \cite{ts} and further 
developped in \cite{bk1} and \cite{fgr}.
More recently  \cite{z1,bk2}, interesting tools for the classification
of N$\times$N matrix QES operators in one spatial dimension have
been constructed and applied to the case N=2 \cite{sz} (although
potentials of the Lam\'e-type are not included in this classification).

Here we consider a suitable class of finite dimensional
vector spaces of N polynomials in a real variable
and we construct families of operators preserving 
sub-classes of these vector spaces.
The corresponding  QES equations respectively 
constitute ``coupled channel'' generalizations of 
the scalar QES equations.

\par Following the basic idea of QES operators \cite{tur1} we
consider the finite dimensional vector space of 
N-uples  of polynomials
of given degree $n_1,n_2,\dots , n_N$
 in a real variable $x$. We slightly generalize
this vector space by setting
\be
\label{espace}
{\cal V} = P \ \left(
{\cal P}(n_1) \oplus  {\cal P}(n_2)\oplus
 \dots \oplus {\cal P}(n_N) \right)
\ee
where ${\cal P}(n_{i})$, $i=1,...N$ denotes the set of real polynomials of degree at
most $n_i$ in $x$ while $P$ is a fixed invertible $N\times N$ matrix
operator; $P$ can be interpreted as a change of basis in the vector
space ${\cal P}(n_1) \oplus \dots \oplus {\cal P}(n_N)$. With such an interpretation, 
and if we assume $n_1 \geq n_2 \geq \cdots \geq n_N$, it is reasonable to 
choose  the matrix $P$ as a lower triangular
matrix with $P_{ii}=1$.

\section{QES Anharmonic matrix potentials}

\vspace{1 cm}

\par We consider an operator of the form
\begin{equation}
\label{hy}
H(y) = -{d^2\over{dy^2}} \unop_N + M_3(y^2)
\end{equation}
where $M_3(x)$ is a $N\times N$ hermitian matrix whose elements are polynomials  of degree at most three in the argument $x$.
After a standard "gauge transformation" of $H(y)$ with the factor
\begin{equation}
\label{hx}
\phi(y) = \exp-\lbrace {p_2\over 2} y^4 + p_1 y^2\rbrace
\end{equation}
($p_1, p_2$ are arbitrary real parameters, $p_2>0$) and the change of variable 
$x=y^2$, the operator equivalent to (\ref{hy}) reads
\begin{eqnarray}
\label{hcha}
\hat H(x) &=& -(4x{d^2\over{dx^2}} + 2{d\over{dx}})\unop_N \nonumber\\
&+& 8 (p_2x+p_1) x{d\over{dx}} \unop_N\nonumber\\
&-& (4 p^2_2 x^3 + 8 p_1p_2x^2+(4p^2_1-6p_2)x-2p_1)\unop_N\nonumber\\
&+&  M_3 (x)
\end{eqnarray}
We now determine the form of the matrix $M_3(x)$ such that
the operator $\hat H(x)$ possesses a finite dimensional invariant vector space of the type (\ref{espace}) for generic values of $p_1, p_2$.
In this purpose, the differential operators in the second line
 of (\ref{hcha}) (i.e. $8 p_2 x^2 d/dx$ and $8 p_1 x d/dx$)
 have separately to be completed into operators which 
have this property.
In order to make use of the results of \cite{z1,bk2} 
we conveniently rewrite  $\hat H$ according to
\begin{eqnarray}
\label{hn}
\hat H(x) &=& -(4x{d^2\over{dx^2}} + 2{d\over{dx}})\unop_N \nonumber\\
          &+& 8 p_2 Q_+ + 8 p_1 Q_0 + 8 p_2(W_3(x) + B)
\end{eqnarray}
where the operators $Q_+ , Q_0$ are defined by
\be
\label{q}
     Q_+ = x^2 \frac{d}{dx} + 2 x A - B \ \ \ , \ \ \ Q_0 = x \frac{d}{dx} + A
\ee
and the  constant matrices $A,B$ are chosen to obey $[A,B] = B$, 
in such a way that
\be
   [Q_+,Q_0] = Q_+   \ .
\ee
Without loosing generality \cite{bk2}, we can  choose 
$A$ and $B$ in the form
\be
\label{ab}
A = {\rm Diag}(0,1,2,\dots , N-1) - \frac{p}{2} \unop_N \ \ \ , \ \ \ 
B_{a,b} = c_b \delta_{a,b+1} 
\ee
The matrix $W_3(x)$ is defined by identification of 
(\ref{hcha}) and (\ref{hn}); it is symmetric and does not contain derivative.
In the following, we further assume this matrix to be 
irreducible by $x$-independent changes of basis; 
in particular, we exclude the cases where $W_3$ is diagonal.

In \cite{z1,bk2}, it was demonstrated that $Q_+, Q_0$
admit a finite-dimensional vector space; we now construct it explicitely.
For this purpose, we define a family of vector spaces
characterized by two integers $N$ and $p$
\begin{equation}
\label{space1}
{\cal V}(N,p) =  {\cal P}(p)\oplus {\cal P}(p-2)\oplus \dots \oplus {\cal P}(p-2N+2)
\end{equation}
and an (invertible) $N \times N$ matrix operator $P$ with matrix elements
\begin{eqnarray}
\label{p}
P_{ij} &=& 0 \quad {\rm if} \quad i<j\nonumber\\
&=& 1 \quad {\rm if} \quad i=j\nonumber\\
&=& \left(\prod^{i-j-1}_{k=0} { c_{j+k} \over{(p+2-2j-k)}}\right)
{1\over{(i-j)!}} ({d\over {dx}})^{i-j}\quad {\rm if} \quad i>j
\end{eqnarray}
The vector space ${\cal V'} \equiv P {\cal V}$ can be seen as 
a change of basis on the space ${\cal V}$ (for brevity, we 
do not write the dependence on $p,N$ anylonger).

The following proposition, which can be checked after an algebra,
provides the explicit form of the invariant vector space of $Q_+, Q_0$~:  
\vspace{0.5cm}

\textsf{Proposition 1}

\vspace{0.5cm}
\begin{eqnarray}
P^{-1} B P &=& P^{-1} (Q_+ + B) P - (Q_+ + B)\\
P^{-1} Q_+ P &=& x^2 \frac{d}{dx} + 2 x A \\
P^{-1} Q_0 P &=& Q_0
\end{eqnarray}
Because the operators on the right hand side of these equalities preserve ${\cal V}$,
it follows immediately that $Q_+$ and $Q_0$ preserve the space ${\cal V'}$.

The requirement that the $(W_3+B)$-part of $\hat H$ also
preserves ${\cal V'}$ is guaranteed by the Proposition 2 below.
For later use we will note by $J_+ , J_0 , J_-$ the usual irreducible representation
of sl(2) by $N \times N$ matrices, i.e.
\be
          [J_0, J_{\pm}] = \pm J_{\pm} \ \ \ , \ \ \ [J_+ , J_-] = 2 J_0 .
\ee
 In particular we can set 
$J_0 = A - \frac{1}{N} {\rm Tr} A$.

\vspace{0.5cm}

\textsf{Proposition 2}

\vspace{0.5cm}

\par If $W_3(x)$ is symmetric and irreducible, then the condition
\be
    P^{-1} (W_3+B) P  {\cal V} \subseteq {\cal V}      
\ee
is fulfilled if and only if
\be
\label{cond}
        B = c J_+ \ \ \ , \ \ \ W_3 = -c(J_+ + J_-)
\ee
The values of the different parameters $c_k$ entering in the matrix $B$
are therefore fixed  up to the multiplicative factor $c$.

So far, we proved that the condition is necessary and sufficient
 for $N$ up to seven and we are confident  that it is true for arbitrary $N$.

The final step is to check whether the operator in the first line of 
(\ref{hn})
also preserves ${\cal V'}$. In this respect, it is sufficient to study the condition
$A \frac{d}{dx} {\cal V'} \subset {\cal V'}$. From the structure of the vector space
${\cal V'}$, it is quickly seen that this holds only for N=2. This shows that
the generalization of the QES anharmonic oscillator to matrices is possible
only for two-dimensional  matrices and the 3-parameter family of operators of \cite{z1} is recovered.

\par Our results thus lead to a family of $3$ parameters matrix 
quasi exactly solvable anharmonic oscillators ($p_1, p_2, c$). 
The parameters $p_1,p_2$ determine the anharmonic
frequencies of the $y^6$- and $y^4$-parts of the potential $M_3$,
 while the constant $c$ ensures a non-trivial 
coupling between the different oscillators.

\section{Clebsch-Gordan coefficients}
The results presented in the  previous section contains a formulation 
of the Clebsch-Gordan matrix of the tensor product
of two representations of sl(2) in terms of differential
operators. In this section we would like to 
further analyze some aspects of this realization.

We consider on the one hand
the standard representation of the algebra sl(2) by 
$N \times N$-matrix generators
noted $J_+ , J_0 , J_-$
and on the other hand the realization expressed in terms of differential
operators~: 
\be
   j_+ = x^2 \frac{d}{dx} + 2 \mu x  \ \ \ , \ \ \
   j_0 = x \frac{d}{dx} + \mu \ \ , \ \ 
   j_- = \frac{d}{dx} 
\ee
 With the choice 
 $\mu = -n/2$ ($n$ integer), this realization preserves the vector space ${\cal{P}}(n)$ and  constitutes the corner stone of QES equations \cite{tur1}.

 Then, we consider the tensor product of these representations, the generators of which are
\be
\label{sum}
        \tilde Q_{\pm,0} =  J_{\pm,0} + j_{\pm,0} \ \ .
\ee
Again, if $\mu = -n/2$ , it
is finite dimensional and preserves, in a reducible way,   
the vector space 
\be
\label{space0}
  {\cal V}_0 = {\cal P}(n) \oplus {\cal P}(n) \dots \oplus {\cal P}(n)
       \ \ \ , \ \ \ N {\rm \ times}
\ee
The explicit decomposition of the representation 
(\ref{sum}), (\ref{space0})
into irreducible ones is achieved in two steps. We define a first 
transformation by means of
\be
\label{tra1}
             Q_{\pm,0} = e^{x J_-} \tilde Q_{\pm,0} e^{-x J_-} \ \ ,
\ee
leading to
\be  
   Q_+ = x^2 \frac{d}{dx} + 2 x(J_0 + \mu \unop) + J_+  \ \ \ , \ \ \
   Q_0 = x \frac{d}{dx} + J_0 + \mu \unop      \ \ , \ \ 
   Q_- = \frac{d}{dx} 
\ee
which can be identified with the operators (\ref{q}) if
$A = J_0 + \mu \unop$, i.e. if $\mu = \frac{N-p-1}{2}$.
In other words the two irreducible representations $J_{\epsilon}$, $j_{\epsilon}$ 
under investigation have dimensions 
$N$ and $p-N+2$, respectively.

The second step leading to the desired decomposition reads 
\be
        \overline Q_{\pm,0} = P^{-1}  Q_{\pm,0} P  
\ee
where the operator $P$ is defined in eq.(\ref{p}). The form of the 
operators $\overline Q$ is available from Proposition 1~:
\be
\overline Q_+ = x^2 \frac{d}{dx} + 2 x  A   \ \ \ , \ \ \
\overline   Q_0 = x \frac{d}{dx} + A \ \ , \ \ 
\overline   Q_- = \frac{d}{dx} 
\ee
It clearly reveals that the vector 
 space (\ref{space1}) is preserved. As a consequence the matrix
$P_{cg} = \exp(-x J_-) P$ constitutes the Clebsch-Gordan matrix
of the decomposition.


It should be noticed that the operators 
$J_-$, $j_-$  preserve separately the vector space (\ref{space0}).
Then, since the transformation (\ref{tra1}) commutes with $J_-$, it results that
both $J_-$ , $j_-$ also preserve ${\cal V'}$.
This statement in fact provides the proof that the condition 
(\ref{cond}) of the Proposition 2 above is indeed sufficient.

\section{Lam\'e-type operators.}

In the following the Jacobi elliptic functions 
\be
{\rm{sn}}(z,k) \quad , \quad {\rm{cn}} (z,k)\quad , \quad {\rm{dn}}(z,k)
\ \ .
\ee
of argument $z$ and modulus $k$ \cite{arscott} are abreviated
respectively by ${\rm{sn}}, {\rm{cn}},{\rm{dn}}$.
These functions are periodic with period $4K(k),
4K(k), 2K(k)$ respectively  ($K(k)$
is the complete elliptic integral of the first type).

It is well known that in order to reveal the  algebraic
properties of the Lam\'e equation
\be
\label{lame1}
       [- \frac{d^2}{dz^2} + k^2 N(N+1) {\rm sn}^2] \psi(z) = E \psi(z) \ \ ,
\ee
the relevant change of variable consists in posing $x = {\rm sn}^2(z,k)$.
In particular the second derivative operator is transformed into
\be
\label{fush}
      {d^2 \over dz^2} = 
4x(1-x)(1-k^2x) {d^2 \over d^2 x} 
+ 2(3 k^2 x^2 - 2 (1+ k^2)x + 1){d \over d x} 
\ee
and Eq. (\ref{lame1})  becomes a Fushs equation  
with four regular-singular points (at $x= 0,1/k^2,1,\infty$).

To our knowledge, attempts to construct (and classify) Schr\"odinger
QES matrix operators with this type of change of variable 
have not been
attempted so far. Particular cases are emphasized in \cite{bk1,by}.
The natural choice is to consider $N\times N$  Schr\"odinger 
matrix operators with potentials depending on the Jacobi ellipitc functions  and
which possess algebraic properties similar to Eq.(\ref{lame1}).
More specifically, we  consider operators of the form
\begin{equation}
\label{lamen}
H(z) = -{d^2\over{dz^2}} \unop_N + V_D(z) + V_I(z)
\end{equation}
with
\begin{equation}
V_D = {\rm sn}^2 {\rm diag} (a_1,a_2,\dots, a_N) + {
\rm diag} (b_1,b_2,\dots,b_N)
\end{equation}
where $a_j, b_j$ denote real constants (without loosing generality
we assume $\sum_{j=1}^N b_j = 0$) and $V_I$ is a symmetric off-diagonal
matrix of the form
\[(V_I)_{ij}=\left\{ \begin{array}{ll}
 \theta_{ij}{\ \rm sn}^{\alpha_{1ij}}
                {\rm cn}^{\alpha_{2ij}}
                 {\rm dn}^{\alpha_{3ij}} & \mbox{if $i \neq j$ } \\
      0   & \mbox{ if  $i = j$}
\end{array}
\right. \] 
Owing the periodicity of the Jacobi elliptic functions, the family
hamiltonian above is to be considered on the Hilbert space of
periodic functions on $[0,4K(k)]$.

Unfortunately, it has not been possible to classify the QES operators
of the form (\ref{lamen}), here we will describe the cases that we were able to  treat.  The difficulty of achieving the classification will 
appear from  these few examples.
The properties of the Jacobi functions that are useful to make  the calculations  are listed in the appendix.

\subsection{case N=1}
This case corresponds to the Lam\'e equation which was discussed
lengthly e.g. in \cite{bgk}.

\subsection{Case N=2}
The non-diagonal potential reduces to only one component
\be
      V_I = \theta \ {\rm sn}^{\alpha_1}
                {\rm cn}^{\alpha_2}
                 {\rm dn}^{\alpha_3} \sigma_1
\ee
Using a similarity transformation of the form
\be
    \hat H(x) = U^{-1}(z) H(z) U(z) \ \ , \ \ 
     U(z) = {\rm diag} ({\rm{sn}}^{\beta_1}\ {\rm{cn}}^{\beta_2}\ {\rm{dn}}^{\beta_3},
{\rm{sn}}^{\gamma_1}\ {\rm{cn}}^{\gamma_2}\ {\rm{dn}}^{\gamma_3})
\ee 
sets the operator (\ref{lamen}) into a form with polynomial coefficients
in the variable $x = {\rm sn}^2$ provided (see Appendix)
\begin{itemize}
\item $\beta_j, \gamma_j = 0 \ {\rm or} \ 1$ \ \ , \ \ $j=1,2,3$
\item $\alpha_j \pm (\beta_j - \gamma_j) =$ non-negative even integer \ \ , \ \ $j=1,2,3$.
\end{itemize}
Considering first the cases where $V_{12}$ is a constant 
($\alpha_{1,2,3}$ = 0)
or a single linear factor in one of the Jacobi functions, we
were able to show that
the only possible QES operators are available for $\theta=0$.

For the three cases a) $V_{12}= \theta {\rm \ sn \ cn}$, 
b) $V_{12}= \theta {\rm \ sn \ dn}$ and c) $V_{12}= \theta {\rm \ cn
\ dn}$,
the construction of non-decoupled QES operators is possible. 
One remarkable feature
is that in each case, two sets of values of the coupling constants
$a_1,a_2,b_1,\theta$ lead to four different algebraizations
of the corresponding operator. We now discuss it in detail
(posing $b\equiv b_1 = - b_2$).

\subsubsection{Case $V_{12} = \theta {\rm \ sn \ cn}$}
The two sets of values of the coupling constants leading to QES
operators are noted Type 1 and Type 2.

\smallbreak
\noindent {\sc{Type 1}}
\smallbreak

\noindent $a_1= k^2(4m^2+2m+1) - 2 b$\\
$a_2 = k^2(4m^2+2m+1) + 2 b$\\
$\theta^2 = 4 b^2 - k^4 (1+ 4m)^2$\\
The parameters $b$ and $k$ remain free,  $m$ is an integer.
Four invariant spaces are available~: 
\begin{eqnarray}
{\cal V}_1 &=&
\left(\begin{array}{cc}
{\rm{sn}} &0\\
0 &{\rm{cn}}
\end{array}\right)
\left(\begin{array}{cc}
1 &\kappa_1\\
0 &1
\end{array}\right)
\left(\begin{array}{c}
{\cal P}(m-1)\\
{\cal P}(m)
\end{array}\right)
\quad , \quad \kappa_1=     \frac{-\theta}{2b+k^2(1+4m)}\\
{\cal V}_2 &=&
\left(\begin{array}{cc}
{\rm{cn}} &0\\
0 &{\rm{sn}}
\end{array}\right)
\left(\begin{array}{cc}
1 &\kappa_2 \\
0 &1
\end{array}\right)
\left(\begin{array}{c}
{\cal P}(m-1)\\
{\cal P}(m)
\end{array}\right)
\quad , \quad \kappa_2=  \frac{\theta}{2b-k^2(1+4m)}\\
{\cal V}_3 &=&
\left(\begin{array}{cc}
{\rm{dn}} &0\\
0 &{\rm{sn}}\ {\rm{cn}}\ {\rm{dn}}
\end{array}\right)
\left(\begin{array}{cc}
1 &\kappa_3 x\\
0 &1
\end{array}\right)
\left(\begin{array}{c}
{\cal P}(m-1)\\
{\cal P}(m-1)
\end{array}\right)
\quad , \quad \kappa_3= - \frac{1}{\kappa_2} \\
{\cal V}_4 &=&
\left(\begin{array}{cc}
{\rm{sn}}\ {\rm{cn}}\ {\rm{dn}} &0\\
0 &{\rm{dn}}
\end{array}\right)
\left(\begin{array}{cc}
1 &0\\
\kappa_4 x &1
\end{array}\right)
\left(\begin{array}{c}
{\cal P}(m-1)\\
{\cal P}(m-1)
\end{array}\right)
\quad , \quad \kappa_4= -\frac{1}{\kappa_1}
\end{eqnarray}

\smallbreak
\noindent{\sc{Type 2}}
\smallbreak

\noindent
$a_1= k^2(4m^2+6m+1)-2b$\\
$a_2= k^2(4m^2+6m+1) +2b$\\
$\theta^2 = 4 b^2 - k^4 (3+4m)^2$\\
 The corresponding invariant vector spaces read

\begin{eqnarray}
{\cal V}_5 &=&
\left(\begin{array}{cc}
1 &0\\
0 &{\rm{sn}}\ {\rm{cn}}
\end{array}\right)
\left(\begin{array}{cc}
1 &\kappa_5 x\\
0 &1
\end{array}\right)
\left(\begin{array}{c}
{\cal P}(m)\\
{\cal P}(m)
\end{array}\right)
\quad , \quad \kappa_5=\frac{k^2(4m+3)-2b}{\theta}\\
{\cal V}_6 &=&
\left(\begin{array}{cc}
{\rm{sn}}\ {\rm{cn}} &0\\
0 &1
\end{array}\right)
\left(\begin{array}{cc}
1 &0\\
\kappa_6 x &1
\end{array}\right)
\left(\begin{array}{c}
{\cal P}(m)\\
{\cal P}(m)
\end{array}\right)
\quad , \kappa_6= \frac{k^2(4m+3)-2b}{\theta}\\
{\cal V}_7 &=&
\left(\begin{array}{cc}
{\rm{sn}}\ {\rm{dn}} &0\\
0 &{\rm{cn}}\ {\rm{dn}}
\end{array}\right)
\left(\begin{array}{cc}
1 &\kappa_7 \\
0 &1
\end{array}\right)
\left(\begin{array}{c}
{\cal P}(m-1)\\
{\cal P}(m)
\end{array}\right)
\quad , \kappa_7 =\frac{2b-k^2(4m+3)}{\theta}\\
{\cal V}_8 &=&
\left(\begin{array}{cc}
{\rm{cn}}\ {\rm{dn}} &0\\
0 &{\rm{sn}}\ {\rm{dn}}
\end{array}\right)
\left(\begin{array}{cc}
1 &\kappa_8 \\
0 &1
\end{array}\right)
\left(\begin{array}{c}
{\cal P}(m-1)\\
{\cal P}(m)
\end{array}\right)
\quad , \kappa_8=\frac{2b+k^2(4m+3)}{\theta}
\end{eqnarray}

\subsubsection{Case $V_{12} = \theta {\rm \ sn \ dn}$}
In the following cases we just write the values of the coupling constants
leading to QES operators; the corresponding invariant vector spaces are similar 
to the ones above, their explicit form can be obtained in a straightforward way.
\smallbreak
\noindent{\sc{Type 1}}
\smallbreak
\noindent
$a_1= k^2(4m^2+2m+1-2b)$\\
$a_2= k^2(4m^2+2m+1+2b)$\\
$\theta^2 = 4 k^2 b^2 - k^2 (1+4m)^2$\\

\smallbreak
\noindent{\sc{Type 2}}
\smallbreak
\noindent
$a_1= k^2(4m^2+6m+1-2b)$\\
$a_2= k^2(4m^2+6m+1+2b)$\\
$\theta^2 = 4 k^2 b^2 - k^2 (3+4m)^2$\\

\subsubsection{Case $V_{12} = \theta  {\rm \ cn \ dn}$}
\smallbreak
\noindent{\sc{Type 1}}
\smallbreak
\noindent
$a_1= k^2(4m^2+2m+1)-2b\frac{k^2}{1+k^2}$\\
$a_2= k^2(4m^2+2m+1) +2b\frac{k^2}{1+k^2}$\\
$\theta^2 =  k^2(1+4m)^2 - 4 b^2 \frac{k^2}{1+k^2}      $\\

\smallbreak
\noindent{\sc{Type 2}}
\smallbreak
\noindent
$a_1= k^2(4m^2+6m+1)-2b\frac{k^2}{1+k^2}$\\
$a_2= k^2(4m^2+6m+1) +2b\frac{k^2}{1+k^2}$\\
$\theta^2 = k^2 (1+4m)^2 - 4 b^2 \frac{k^2}{1+k^2}$\\

Concerning the operators, we want to point out the following
things:
\begin{itemize}
\item The extraction of the prefactor is done in two steps. After factorizing the
appropriate products of Jacobi functions, the non-diagonal part of the potential
takes the form
\be
\label{matrix}
\left(\begin{array}{cc}
 0 &x^{2-\epsilon}\\
 x^{\epsilon} &0
\end{array}\right)
\quad , \quad  \epsilon = 2 \ {\rm or} \ 1 \ {\rm or} \ 0
\ee
which is clearly incompatible with an operator preserving a vector space
of the form 
${\cal P}(n) \oplus {\cal P}(m)$ for integers $m,n$. The setting of the operator
in a form preserving such a vector space is realized by the second prefactor
(the triangular matrix one). As an example if $\epsilon=0$ in 
(\ref{matrix}), then
use is made of the relations
\be
\left(\begin{array}{cc}
 1 &-A\\
 0 &1
\end{array}\right)
\left(\begin{array}{cc}
 0 &1\\
 0 &0
\end{array}\right)
\left(\begin{array}{cc}
 1 &A\\
 0 &1
\end{array}\right)   =
\left(\begin{array}{cc}
 0 &1\\
 0 &0
\end{array}\right)
\ee
\be
\left(\begin{array}{cc}
 1 &-A\\
 0 &1
\end{array}\right)
\left(\begin{array}{cc}
 0 &0\\
 1 &0
\end{array}\right)
\left(\begin{array}{cc}
 1 &A\\
 0 &1
\end{array}\right)   =
\left(\begin{array}{cc}
 -A &A^2\\
 1 &A
\end{array}\right)
\ee
with $A \div x$ in order to eliminate the quadratic power of $x$ occuring in $V_{12}$.
\item For $V_{12} = \theta {\rm \ sn \ cn}$, the limit $k\rightarrow 0$ is non trivial
and leads to a potential of the form
\be
\left(\begin{array}{cc}
 {\rm cos}^2(x)- \frac{1}{2} &{\rm cos}(x) {\rm sin}(x)\\
 {\rm cos}(x) {\rm sin}(x)&{\rm sin}^2(x)- \frac{1}{2}\end{array}\right)
\ee
In particular, the dependence on  $m$ disappears, so this integer can be choosen
arbitrarily large and an infinite set
of algebraic eigenvectors occurs in this limit.
This confirms the fact that the
corresponding Schr\"odinger equation is completely solvable and is related to the
stability of some soliton solutions occuring in the Goldstone model in 1+1 dimensions
\cite{bt} with a periodic condition for the space coordinate.

\item The case $V_{12} = \theta {\rm \ cn \ dn}$ with $b=(1+k^2)/2$ was 
studied lengthly in \cite{bk1}.
The corresponding Schr\"odinger equation is related to the stability analysis of the sphaleron
solution in the U(1)-Abelian Higgs model in 1+1 dimensions (again with a periodic 
condition for the space coordinate).
\end{itemize}

\subsection{Case N=3}
 We attempted to construct
3$\times$3 matrix operators with some choice for the constants
$\alpha_{a i j}$ ($a,i,j = 1,2,3$) inspired from the results
above. Several trials were unsuccesfull
(namely with $V_{12} \div V_{23} \div {\rm sn \ cn}$);
however  the choice
\be
\label{pot3}
 V_I =
\left(\begin{array}{ccc}
0             &\theta_1 \ {\rm cn \ dn}         &V_{23}  \\
\theta_1 \ {\rm cn \ dn}     &0 &\theta_2 \ {\rm cn \ dn} \\
V_{23}& \theta_2 \ {\rm cn \ dn}  &0
\end{array}\right)\
\ee
leads to the wanted form of equations 
and preserves a vector space of the form
\begin{equation}
\label{espace3}
{\cal V} =
\left(\begin{array}{ccc}
{\rm cn} &0&0\\
0 &{\rm dn}&0\\
0&0&{\rm cn}
\end{array}\right)\  . \
\left(\begin{array}{ccc}
1 &\alpha&\beta\\
0 &1&\gamma\\
0&0&1
\end{array}\right)\ .  \ 
\left(\begin{array}{c}
{\cal P}(n-2)\\
{\cal P}(n-1)\\
{\cal P}(n)
\end{array}\right)\qquad \qquad
\end{equation}
The condition
\begin{equation}
H(z) {\cal V} \ \subseteq {\cal V}
\end{equation}
leads to ten equations for the parameters $\alpha, \beta, \gamma, k, a_i, b_i$ and $\theta_a$.
These equations are compatible with each other only if $k=1$ and 
$V_{23} = \theta_3 {\rm cn^2}$.
 We found it convenient to leave the parameters $\alpha, \beta, \gamma$ free
and to express the coupling constants $a_i, b_i$ and $\theta_a$ in terms of them.
We refrain to write these tedious expressions in general but we mention that they have the polynomial
\begin{equation}
(1+\gamma^2+\beta^2) (1+\beta^2+\gamma^2+\alpha^2\gamma^2-2\alpha\beta\gamma)
\end{equation}
as common denominator. Also they are such that
\begin{equation}
a_1+a_2+a_3 = 2(6n^2-3n+4)
\end{equation}
For the case with e.g. $\alpha = \beta = \gamma = 1$ we find
\begin{equation}
a_1=a_2 = {12n^2-10n+11\over 3}\quad , \quad a_3 = {2\over 3} (6n^2+n+1)
\end{equation}
\begin{equation}
b_1=   b_2 =  {4n-3\over 3} \quad , \quad  b_3 = -2 b_1
\end{equation}
\begin{equation}
\theta_1 = {7-2n\over 6} \quad , \quad 
\theta_2=- {4n+1\over 3}\quad , \quad  
\theta_3 = {2(4n+1)\over 3}
\end{equation}
as another example, if $\alpha = - \beta = 1 \ , \ \gamma = 0$, we have
\begin{equation}
a_1= {2(6n^2-7n+3)\over 3}\quad , \quad a_2=a_3 = {1\over 3} (12n^2-2n+9)
\end{equation}
\begin{equation}
b_1= \frac{2(4n+1)}{3}\quad , \quad b_2 = -{4n+1\over 3} \quad , \quad
b_3 = -b_1-b_2
\end{equation}
\begin{equation}
\theta_1 = {3-4n\over 3} \quad , \quad 
\theta_2={20n-3\over 6}\quad , \quad 
\theta_3 = {2(4n-1)\over 3}
\end{equation}

Remarkably, replacing  the first factor in (\ref{espace3}) by
${\rm diag (dn,cn,dn)}$, we obtain the same solution for $a_1,a_2,
\dots$. So the operator (\ref{pot3}) possesses at least a double
algebraization, but we have not attempted to construct the other
possibly existing ones yet.

The 3$\times$3-matrix QES potential (\ref{pot3}) 
strongly contrasts with the 2$\times$2 ones obtained above.
While the former depends on three free parameters but  
exists only on the full line
(since $k=1$), the latter
can have an arbitrary period (equal to $4K(k)$) but
has only one free parameter 
(for instance noted by $b$)  for
a fixed $k$. 
The fact that the operator $H(z)$ for
a 3$\times$3-matrix QES potential 
is quasi exactly solvable only for $k^2=1$ was unexpected to us. 
Accordingly, this operator corresponds to a $3\times 3$ 
matrix version of the P\"oschl-Teller operator.

\section{Concluding remarks}
In the first part of this paper we obtained the matrix
generalization
of the celebrated sextic QES anharmonic oscillator.
The examples of operators presented in the second part 
reflect the difficulty to classify the coupled-channel
(or matrix) QES Schr\"odinger equations when the change of variable
involves several singular points like in (\ref{fush}). 
We hope that this note will  motivate further investigation
of this problem.

\newpage

\section {Appendix A}
For $k=0$ we have $K(0)= \frac{\pi}{2}$
and the Jacobi functions reduce to standard trigonometric functions~:
${\rm sn}(z,0)= {\rm sin}(z)$, ${\rm cn}(z,0) = {\rm cos}(z)$, ${\rm dn}(z,0)=1$.
For $k=1$, we have $K(1) = \infty$
and the Jacobi functions reduce to elementary functions~:
\begin{equation}
{\rm sn}(z,1) = {\rm tanh} (z)
\end{equation}
\begin{equation}
{\rm cn}(z,1) = {\rm dn} (z,1) = {1\over {{\rm cosh}(z)}}
\end{equation}
In this limit, the Lam\'e equation
becomes a Poschl-Teller equation.
For generic values of $k$, the Jacobi
functions obey the following relations~: 
\be
{\rm{cn}}^2+{\rm{sn}}^2=1\quad , \quad {\rm{dn}}^2+k^2{\rm{sn}}^2=1
\ee
\be
{d\over{dz}}{\rm{sn}} = {\rm{cn}}\  {\rm{dn}}\quad ,
\quad {d\over{dz}} {\rm{cn}} = -{\rm{sn}}\  {\rm{dn}}\quad
, \quad {d\over{dz}} {\rm{dn}} = -k^2 {\rm{sn}}\  {\rm{cn}}
\ee
These identities as well as the following ones 
are useful to establish the equations in the variable
$x= {\rm sn}^2$ after
the prefactor including the Jacobi functions has been extracted~:
\begin{eqnarray}
f           &f''/f         &({\rm sn \ cn \ dn})\ f'/f      \nonumber \\
1            &0              &0                             \nonumber \\
{\rm sn}      &2k^2x-(1+k^2)   &k^2 x^2 - (1+ k^2)x+1        \nonumber \\ 
{\rm cn}      &2k^2x-1         &k^2 x^2 - x                  \nonumber \\    
{\rm dn}      &2k^2x-k^2       &k^2 x^2 - k^2 x              \nonumber \\   
{\rm cn \ dn} &6k^2x-(1+k^2)   &2k^2 x^2 - (1+ k^2)x         \nonumber \\   
{\rm sn \ dn} &6k^2x-(1+4k^2)  &2k^2 x^2 - (1+2k^2)x+1        \nonumber \\   
{\rm sn \ cn} &6k^2x-(4+k^2)   &2k^2 x^2 - (2+ k^2)x+1        \nonumber \\
{\rm sn \ cn \ dn} &12k^2x-4(1+k^2)  &3k^2 x^2 - 2(1+ k^2)x+1           \nonumber
\end{eqnarray}
No similar identities are available (to our knowledge) with 
different choices of the function $f$.

\end{document}